# Machine Learning Framework for Quantum Sampling of Highly-Constrained, Continuous Optimization Problems

Blake A. Wilson[1,3,*], Zhaxylyk A. Kudyshev[1,3,*], Alexander V. Kildishev[1], Sabre Kais[2,3], Vladimir M. Shalaev[1,3], and Alexandra Boltasseva[1,3]

[1]School of Electrical and Computer Engineering, Birck Nanotechnology Center and Purdue Quantum Science and Engineering Institute, Purdue University, West Lafayette, IN, USA

[2]School of Chemistry, Purdue University, West Lafayette, IN 47907, USA

[3]The Quantum Science Center (QSC), a National Quantum Information Science Research Center of the U.S. Department of Energy (DOE), Oak Ridge, TN 37931

[*]authors with equal contribution

**Abstract**

In the recent years, there is a growing interest in using quantum computers for solving combinatorial optimization problems. In this work, we developed a generic, machine learning-based framework for mapping continuous-space inverse design problems into surrogate quadratic unconstrained binary optimization (QUBO) problems by employing a binary variational autoencoder and a factorization machine. The factorization machine is trained as a low-dimensional, binary surrogate model for the continuous design space and sampled using various QUBO samplers. Using the D-Wave Advantage hybrid sampler and simulated annealing, we demonstrate that by repeated resampling and retraining of the factorization machine, our framework finds designs that exhibit figures of merit exceeding those of its training set. We showcase the framework's performance on two inverse design problems by optimizing (i) thermal emitter topologies for thermophotovoltaic applications and (ii) diffractive meta-gratings for highly efficient beam steering. This technique can be further scaled to leverage future developments in quantum optimization to solve advanced inverse design problems for science and engineering applications.

**Introduction**

Combinatorial optimization has recently seen tremendous progress with new algorithms and heuristics, such as simulated annealing, genetic algorithms, and adiabatic optimization[1]. Specifically, the quadratic unconstrained binary optimization (QUBO) formalism of combinatorial optimization has attracted significant interest due to its applicability to a broad range of physical and NP-hard (i.e., non-deterministic polynomial-time hard) combinatorial optimization problems[2–4]. For example, it has been demonstrated that QUBO can be used for factoring integers[5], electronic structure calculations[6], capital budgeting[7], solving the maximum cut problem[8,9], graph coloring[10], traffic flow optimization[11], number partitioning[12], etc. Another key aspect that boosted interest in the QUBO formalism is its isomorphism to Ising Hamiltonians, commonly used in physics and chemistry[13]. This equivalence enables direct mapping of a broad range of physics/chemistry optimization problems into the combinatorial optimization domain and the use of various physical platforms to perform highly efficient QUBO-based optimization via physical processes[14–16].

Recent progress in the development of various near-term quantum computing platforms opens up more efficient ways for addressing the aforementioned optimization problems in terms of time and computational resource requirements by leveraging the power of physical mechanisms, specifically, quantum mechanics, in the processing unit. For example, the D-Wave's quantum annealers are actively used for addressing the QUBO problems via encoding the QUBO parameters into a system of coupled superconducting qubits and retrieving the lowest energy configuration via quantum annealing[17,18]. Surmounting evidence is showing that quantum annealing offers a so-called quantum speedup over classical QUBO sampling methods[19,20]. The QUBO-based optimization consists of three main steps: (i) reformulating the optimization problem into a QUBO model; (ii) embedding the retrieved QUBO model parameters into the QUBO sampler; and (iii) retrieving the global solution of the problem. In most cases, the first step in QUBO-based optimization is realized by exploiting a one-to-one correspondence between the combinatorial problem under consideration and the architecture of the QUBO-solver[21,22]. On the one hand, this correspondence makes retrieving corresponding QUBO parameters of the problem trivial. On the other hand, it significantly reduces the types of problems considered, especially those outside the combinatorial domain.

Another important subfield of optimization problems is continuous optimization. Continuous optimization is built on the continuous domain, real-space parameters with differential, calculus-based relationships. Some simple continuous optimization problems can be solved analytically. However, many problems do not have analytic solutions and are solved numerically by employing search algorithms such as stochastic gradient descent and various heuristics, which cannot guarantee optimality. Many problems exist for which these search algorithms do not work well because of the vast continuous domain. Novel techniques that enable invertible mapping from continuous space optimization problems to QUBO problems may provide a way to take advantage of recent and future advancements in QUBO, including quantum optimization algorithms. Therefore, there is an apparent demand for a universal method of mapping continuous optimization problems into the QUBO formalism to generate better continuous space solutions.

Within this work, we developed a novel machine-learning assisted framework that maps a broad range of continuous optimization problems into QUBO problems and samples their optimized solutions using any available classical or quantum QUBO solver. Specifically, we demonstrated a binary variational autoencoder (bVAE) assisted QUBO framework (bVAE-QUBO) that encodes a continuous optimization problem into a binary, compressed space and samples this compressed space with quantum-assisted QUBO samplers. We showcase the performance of the developed framework on two practical, continuous optimization problems of nanophotonics: (1) optimization of a free-form thermal emitter for thermophotovoltaics (TPV) and (2) optimization of dielectric, free-form diffractive meta-grating for beam steering. Although the developed technique is showcased on inverse design problems of nanophotonics, it can be directly applied to a broad range of practical continuous optimization problems, e.g., in mechanical engineering, chemistry, material synthesis. By employing a quantum-assisted algorithm for continuous optimization, our framework provides a long-sought-after example of a quantum-assisted, machine learning algorithm that has potential for quantum-supremacy[23], uses noisy intermediate-scale quantum platforms for practical engineering problems[24], and could fully leverage future developments in quantum and classical QUBO sampling[25,26].

## Methods

### bVAE-QUBO General Framework

Motivated by the increasing number of qubits in D-Wave quantum annealers[14,17,18] and the recent work by Hastings in proving a relativized speedup for stoquastic adiabatic quantum computing[27], we developed a framework to map highly constrained continuous optimization problems into the QUBO model, which can be minimized by quantum annealers and other QUBO samplers. The first step of the developed technique is compressing a problem dataset from a discretized, continuous optimization problem onto a binary, compressed space by training a binary variational autoencoder (bVAE). We then map the binary, compressed space into an equivalent QUBO problem by training a second-order factorization machine and retrieving corresponding parameters of the QUBO model[28]. Finally, we optimize the retrieved QUBO problem to find an optimal binary vector via a QUBO sampler like the D-Wave quantum annealer. The factorization machine is retrained and resampled repeatedly using a QUBO sampler until it converges to produce good, continuous space solutions. Below we highlight each step of the process in more detail (Fig. 1.).

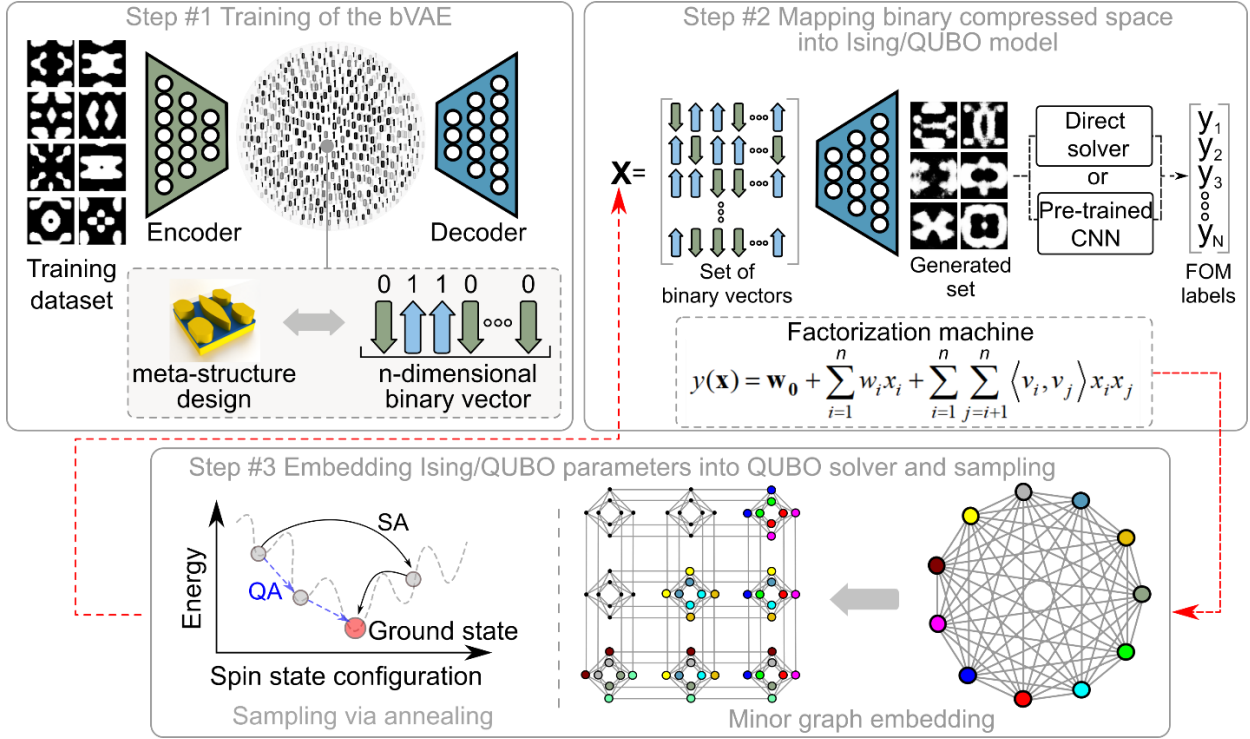

Figure 1. The developed general bVAE-QUBO framework steps: (Step 1) training of the binary variational autoencoder (bVAE) and construction of binary compressed space representation of the dataset (inset); (Step 2) mapping the resulting binary compressed space into the QUBO/Ising model via training of the second order factorization machine; (Step 3) embedding the retrieved Ising/QUBO model parameters into the hardware and sampling the optimized data from the QUBO sampler.

**Step #1** of the developed framework is to train the bVAE network and construct a binary, compressed representation of the optimization problem. This step maps the continuous optimization problem onto the binary domain and substantially reduces the dimension of the continuous space problem. Conventionally used variational autoencoders, which consist of two coupled neural networks (encoder and decoder), construct an invertible mapping between a lower-dimensional encoding of a dataset to the continuous-space solution[29,30]. Recently, it has been demonstrated that variational autoencoders can be generalized for categorical compression of complex 1D and 2D datasets, where the compressed space coordinates are discrete variable vectors[31,32]. Here, we use a limiting case of such categorical variational autoencoders – binary variational autoencoder, which allows the construction of binary, compressed space representation of the complex datasets[33]. A properly trained bVAE can be considered an invertible map, $g : \{0,1\}^n \to \mathbb{R}^{m \times m}$, between the binary, compressed space vectors of size $n$ and the discretized, continuous space solutions of dimensionality $m \times m$. The bVAE network is trained by minimizing a reconstruction loss and the Kullback-Leibler divergence loss. The latter defines the deviation of the recognition distribution (obtained with the model data) from the pre-defined prior. The main difference between the bVAEs and the vanilla variational autoencoders is that the prior and recognition models are under different distributions. The bVAE is under a Bernoulli distribution, while the typical variational autoencoder's distribution is assumed to be Gaussian. We used the Gumbel-Softmax re-parameterization trick[31] for backpropagating the error during training, enabling the derivative calculation on the stochastic nodes of the bVAE network. More details on the bVAE structure and the training process can be found in Supplementary Materials, Section S1.

**Step #2** of the bVAE-QUBO framework maps the bVAE's compressed space into a QUBO problem or Ising Hamiltonian via training a second-order factorization machine. This step exploits the fact that second-order factorization machines are isomorphic to QUBO objective functions and Ising Hamiltonians. Factorization machines, introduced by Rendle for learning sparse feature interactions[28], are low-capacity models that infer coupling coefficients, $\langle v_i, v_j \rangle$, by a factorization matrix, $V \in \mathbb{R}^{n \times k}$. The coupling coefficients, $\langle v_i, v_j \rangle$, are determined by taking the dot product of the $i^{th}$ and $j^{th}$ rows in $V$, which is equivalent to multiplying the factorization matrix by its transpose, $VV^T$. A factorization machine acts on an input binary vector, $x \in \{0,1\}^n$, and returns a figure of merit $y \in \mathbb{R}$.

$$y(x) = w_0 + \sum_{i=1}^{n} w_i x_i + \sum_{i=1}^{n} \sum_{j=i+1}^{n} \langle v_i, v_j \rangle x_i x_j, \tag{1}$$

$w_0 \in \mathbb{R}$ is a global bias, $w \in \mathbb{R}^n$ defines the weights for the discrete components of $x$. All free parameters, $w_0, w$, and $\langle v_i, v_j \rangle$ are optimized via supervised training of the factorization machine. Specifically, the factorization machine is trained on a randomly sampled dataset of binary vectors $X$ and their corresponding figure of merit labels $Y$ from the binary, latent space of the bVAE. The figure of merit labels are calculated by passing a binary vector, $x \in X$, through the bVAE's decoder and calculating the figure of merit on its continuous-space solution.

A crucial benefit to restricting the factorization machine to a second-order model is that QUBO objective functions are only second-order polynomials. A QUBO sampler finds the minimum input binary vector to a second-order pseudo-boolean function via classical or quantum sampling algorithms.

$$\underset{x \in \{0,1\}^n}{\text{argmin}} \sum_{i=0}^{n} Q_i x_i + \sum_{i_1=0}^{n} \sum_{i_2 > i_1} Q_{i_1, i_2} x_{i_1} x_{i_2}, \tag{2}$$

here $x \in \{0,1\}^n$ and $Q$ is a $n \times n$ matrix containing local biases (diagonal terms) and coupling coefficients (off-diagonal terms). An alternative formulation of QUBO problems is the formulation via the Ising Hamiltonian. The restricted Ising Hamiltonian $H(\boldsymbol{\sigma})$ used by quantum annealers contains local biases $\boldsymbol{h}$ and a quadratic term proportional to a qubit coupling matrix $J$.

$$H(\boldsymbol{\sigma}) = \sum_{i=1}^{N} h_i \sigma_i^{(z)} + \sum_{\langle i,j \rangle}^{N} J_{ij} \sigma_i^{(z)} \sigma_j^{(z)}, \qquad (3)$$

$\sigma_i^{(z)}$ is a Pauli-Z matrix acting on the $i^{th}$ qubit's spin state. The spin state must collapse to an eigenvalue of -1 or +1 upon measurement. So, one can loosely consider this collapsed measurement value as a spin variable, $s_i \in \{-1, +1\}$. Quantum annealers employ the adiabatic algorithm to find the minimum input spin state, $\boldsymbol{s}$, for (3) by encoding the magnetic field strengths of superconducting qubits with the local biases $\boldsymbol{h}$ and the coupling matrix $J$. Comparing (1), (2), and (3), we can see a similarity between the parameters of the Ising Hamiltonian, factorization machine, and QUBO objective function:

$$h_i \leftrightarrow w_i \leftrightarrow Q_i, \qquad J_{ij} \leftrightarrow \langle v_i, v_j \rangle \leftrightarrow Q_{i,j} \qquad (4)$$

All three of these models are equivalent for optimization under a simple change of basis between the spin domain of the Ising Hamiltonian and the Boolean domain of the QUBO, see Supplementary materials, S1. Hence, by training the factorization machine regression model and retrieving $\boldsymbol{w_0}, \boldsymbol{w}$, and coupling matrix $\langle \boldsymbol{v_i}, \boldsymbol{v_j} \rangle$, we can create an equivalent Ising or QUBO model for a classical or quantum QUBO sampler.

**Step #3** embeds the retrieved QUBO/Ising problem from the factorization machine in Step #2 into a sampler and retrieves optimized solutions. While this step depends on the type and architecture of the QUBO sampler, we used the D-Wave quantum annealers in our work. The virtual QUBO from the factorization machine is fully connected, meaning each bit has a pairwise coupling coefficient with every other bit, i.e., $\forall i, j \in \{1,2,\ldots,N\}: Q_{i,j} \neq 0$. However, to physically realize a quantum annealer, the physical Ising model cannot be fully connected, i.e., $\exists i, j \in \{1,2,\ldots,N\}: J_{i,j} = 0$.

There are two available Ising connectivity topologies for the D-Wave quantum annealers: Pegasus[34] and Chimera[35]. By using minor-embedding techniques, one can convert any virtual QUBO into any physical Ising model at the expense of adding auxiliary variables. The D-Wave architectures have a fundamental limit in the number of fully connected qubit connections realized by these minor-embedding techniques. This limit is known as the 'clique' size of the Ising Hamiltonian's connectivity graph. The D-Wave Advantage quantum annealer has a Pegasus connectivity graph with a maximum clique size of 180 qubits, while the D-Wave 2000Q's Chimera graph clique size is 64 qubits. Therefore, the factorization machine's input vector size cannot be larger than 180 bits if one wants to use a quantum annealer. Fortunately, the D-Wave also provides simulated annealers and quantum-classical hybrid samplers, which can handle any clique size. While the quantum annealer may ensure asymptotic speedup over classical computers with respect to the QUBO size, simulated annealing and other classical QUBO samplers are not limited to connectivity requirements or maximum clique sizes. Nevertheless, we include more details on these topics in the Supplementary Materials, Section S1.

Once the factorization machine's parameters are embedded into the QUBO sampler, the sampler will return a set of optimized binary vectors, $x_{new}$. The corresponding figure of merit labels, $y_{new}$, for the sampled vectors are assessed by generating their continuous-space solutions via the bVAE's decoder and retrieving the corresponding figure of merit values through a direct solver. Finally, the $(X, Y)$ set initially used for training the factorization machine during the previous iteration is updated with $(x_{new}, y_{new})$ by appending the new vectors to the dataset. Step 3 concludes one iteration of the bVAE-QUBO, while the next iteration starts with retraining the factorization machine on the updated set $(X \cup x_{new}, Y \cup y_{new})$. Steps #2 and #3 are looped until the figure of merit of the sampled solutions reaches saturation or the desired number of iterations is achieved. The main idea is that retraining the factorization machine on sampled vectors increases the variance of the model and forces it to be a better surrogate model for the continuous solution space. Sampling the newly trained factorization machine should not give any previous binary solution unless the sample has a high enough figure of merit, indicating a saturated factorization machine. In the next section, we showcase the performance of the developed bVAE-QUBO framework on the optimization of meta-structure designs for nanophotonic applications.

## Results

### bVAE-QUBO for Nanophotonic Inverse Design Problems

Recently, new optimization frameworks, such as topology optimization[36–40] and metaheuristics[41,42], for nanophotonics have emerged as powerful design algorithms. However, these techniques require significant computational resources and have exponential asymptotic complexity with respect to problem constraints and the dimensions of the optimization parametric space. In response, various machine learning and deep learning algorithms have been adapted to address optimization problems in nanophotonics[43–52]. For example, generative adversarial networks[53,54] and adversarial autoencoders coupled with adjoint topology optimization techniques for optimizing meta-structures produced nonintuitive designs. It was also demonstrated that adversarial autoencoder-based optimization frameworks coupled with metaheuristic algorithms could perform global optimization searches within the compressed design spaces of a pre-trained adversarial autoencoder network[55,56].

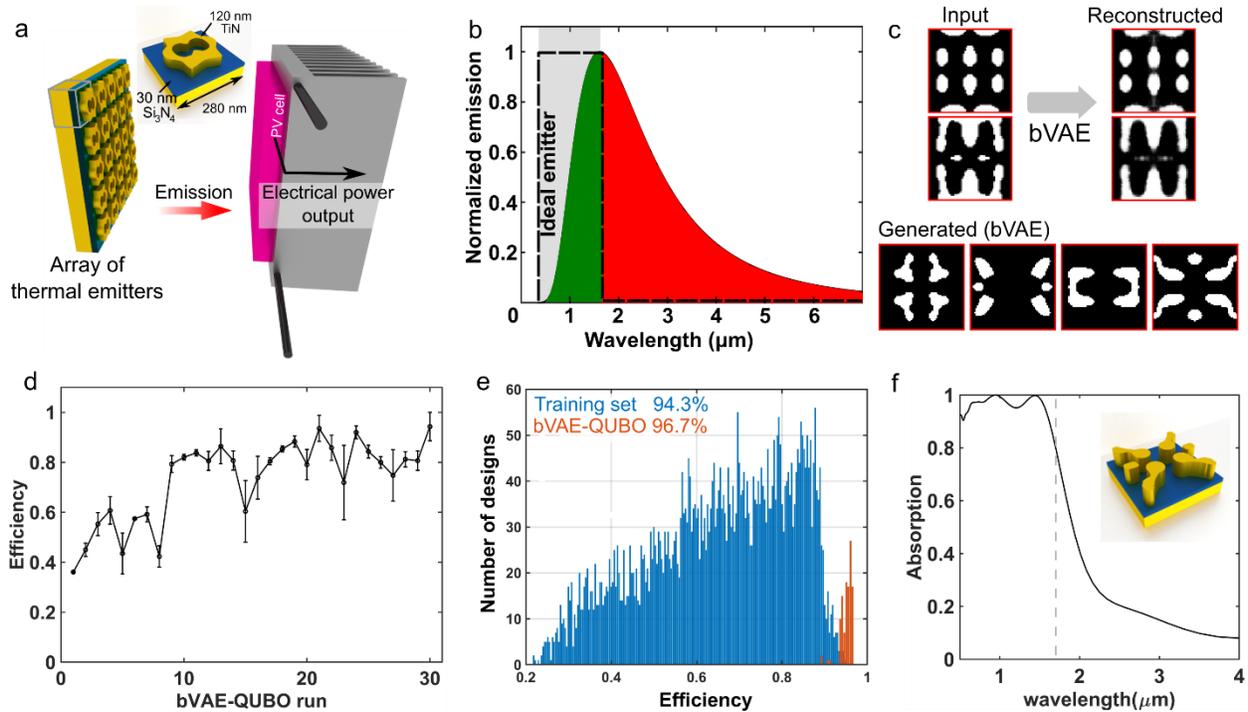

**Figure 2. Simulated annealing assisted bVAE-QUBO for thermal emitter design optimization.** (a) Schematic of a thermophotovoltaic engine: a heater patterned with a thermal emitter array and a photovoltaic cell. Inset shows the base structure under consideration consists of a 300-nm-thick TiN back reflector, 30-nm-thick $Si_3N_4$ dielectric spacer, and 120-nm-thick top TiN patterned layer in 280×280 $nm^2$ unit cell. The top TiN layer is set to be the optimization region. (b) Blackbody radiation of the bare heater (solid black curve) corresponding to emission of blackbody at 1800 °C. The grey rectangular region highlights the GaSb photovoltaic cell working band. Only in-band radiation is converted into electrical power (green area), while out-of-band radiation cases heating of the photovoltaic cell (red area). Black dashed contour corresponds to an ideal thermal emitter's emissivity/absorption spectrum. (c) Two examples of input and reconstructed emitter designs by the trained bVAE network (top) and examples of randomly sampled thermal emitters (bottom). Patterns are top view of the optimization area, white color corresponds to TiN, while black corresponds to air. (d) Convergence plot of 30 iterations of the bVAE-QUBO framework. (e) Efficiency distribution of the training set used for bVAE training (5000 designs, blue histogram) and 100 designs generated via bVAE-QUBO (orange). (f) Emissivity/absorption spectra of the best design sampled with bVAE-QUBO. Vertical dashed line shows the upper bound of the GaSb photovoltaic's working band. Inset shows the schematic of the thermal emitter design.

However, due to the general complexity of the inverse design problems, such approaches may not be effective in the case of highly constrained problems, which demand multi-objective optimization within the high-dimensional latent spaces.

Within this section, we show empirical evidence that the developed bVAE-QUBO framework can address the aforementioned challenges. Specifically, we demonstrate that by using the bVAE-QUBO framework, it is possible to construct a binary, compressed space representation of meta-devices with complex shapes and topologies and map it into a QUBO sampler for optimized, free-form design sampling. In the remainder of this section, we show the results from applying our framework to two case studies of optimizing (i) thermal emitters for thermophotovoltaics and (ii) dielectric, free-form diffractive grating for beam steering.

**Thermal Emitter for TPV Application.**

The TPV engine generates electrical power via radiative heat transfer between a heater and an array of photovoltaic cells (Fig. 2a). High-efficiency power generation in a TPV system requires maximizing the portion of the emission that overlaps with the working band of the photovoltaic cell (in-band radiation, green area in Fig. 2b) and minimizing the rest of the spectra (out-of-band radiation, red area in Fig. 2b)[57–59]. There are three main requirements for implementing a high-efficiency TPV engine: (i) high temperature of the heater (>1000 °C), (ii) refractory material platform for the elements of the TPV system to ensure stable performance of the device at high temperatures, and (iii) pre-optimized emissivity properties of the heater. While the first two constraints can be addressed by choosing a suitable refractory material platform[60–62], the third requirement can be fulfilled by patterning the surface with a properly designed thermal emitting metasurface. In the ideal case, the emissivity should completely overlap the working band of the photovoltaic cell (black dotted step-function, Fig. 2b). Such surface emissivity ensures total cancelation of the parasitic out-of-band radiation, which leads to the reduction of the photovoltaic efficiency due to the heating while maintaining the maximum possible free-carrier generation rate. We consider TPV systems utilizing GaSb photovoltaic cells with a working band ranging from $\lambda_{min}$ to $\lambda_{max}$ (shaded area in Fig. 2b).

Within this work, we consider a gap plasmon metasurface[63,64] configuration consisting of an optically thick back titanium nitride (TiN) reflector, a 30-nm-thick silicon nitride ($Si_3N_4$) spacer, and a 120-nm-thick top layer (optimization region), with a fixed 280-nm lateral periodicity (inset, Fig. 2a). The main goal of the optimization is to determine the topological shape of the material distribution (TiN and air) within the optimization region, which ensures spectral emissivity matching the emissivity of the ideal emitter. For quantification of the device performance, we define the efficiency of the thermal emitter as a product of in-band ($eff^{in}$) and out-of-band efficiencies ($eff^{out}$). $eff^{in}$ is an in-band radiance of the emitter normalized to the in-band radiance of ideal emitter at 1800 C, while out-of-band efficiency $eff^{out}$ is defined as a ratio of the out-of-band radiance of the back reflector and radiance of the optimized design.

The first step of bVAE-QUBO is realized by training the bVAE network on topology optimized thermal emitter designs. The training set consists of 5000 topology optimized designs obtained via a previously developed adversarial autoencoder-based optimization framework. A VGGnet regression model trained on the same dataset for rapidly estimating the thermal emitter's efficiency based on its design. More information on training set generation is in Supplementary materials, S2. During the training of the bVAE, the encoder takes 64 × 64 binary, greyscale topology images (top view of the antenna, Fig. 2c) as an input and trains to compress it into the 500-dimensional binary space. Likewise, the decoder trains to reconstruct the topology of the antenna design based on an inputted 500-dimensional binary vector. After training, the encoder can compress antenna designs into the compressed latent space while the decoder can act as a generator that maps latent vectors to novel meta-structures. Two examples of the reconstructed antenna designs are shown in Fig. 2c. Here one can see that the bVAE network ensures precise reconstruction of complicated antenna designs. Additionally, Fig. 2c shows some examples of the randomly sampled antenna designs using the bVAE's decoder. Note that the Gaussian filtering with 20 nm blur size is applied to the generated patterns to eliminate small features introduced by the bVAE noise.

The second step of the bVAE-QUBO framework starts by training the factorization machine on the binary vectors ($X$) and efficiency labels ($Y$) generated by the bVAE network. This data set is constructed by randomly sampling 11250 binary vectors from the binary, compressed space and calculating their corresponding efficiency labels. The efficiency labels are retrieved by generating their thermal emitter design using the decoder and calculating their efficiency using a pre-trained VGGnet. The ($X, Y$) set is constrained such that half of it corresponds to low-efficiency designs (70-80% efficiency), 30% of the designs in the set have moderate efficiencies (between 80% to 90% ), and 20% of them have more than 90% efficiency. The supervised training of the factorization machine is done using the adaptive gradient descent optimization with the mean square error loss function. 70% of the ($X, Y$) dataset is used for training, while 10% for validation and 20% is used for testing. The trained factorization machine ensures $r^2 = 72\%$ and mean square error of 0.001. Additional information on the structure of the bVAE network and training the factorization machine can be found in Supplementary materials, S1.

Simulated Annealing Assisted bVAE-QUBO Framework

Using simulated annealing as the QUBO sampler, the bVAE-QUBO framework executed 30 iterations. Thermal emitter design efficiencies sampled during each of the bVAE-QUBO runs are shown in Fig. 2d. The data points and error bars show the mean efficiencies and corresponding standard deviations of 10 designs sampled during each of the bVAE-QUBO runs. We can see that updating QUBO parameters via retraining the factorization machine on the newly sampled vectors will significantly increase the quality of sampled designs (from ~40% of the initially trained factorization machine up to >90%). Figure 2e shows the efficiency distributions of the dataset used for bVAE training (5000 designs, blue bars) and the best 100 designs sampled with bVAE-QUBO (orange bars). Finite difference time domain analysis (Lumerical FDTD) is used to assess the final efficiencies of each of the sets after running the framework. The best design in the training set ensures 94.3% efficiency, while one sampled via the simulated annealing-based bVAE-QUBO framework approach ensures 96.7%. The emissivity spectra of the best design in the bVAE-QUBO set are shown in Fig. 2f, while the inset shows the corresponding design of the thermal emitter.

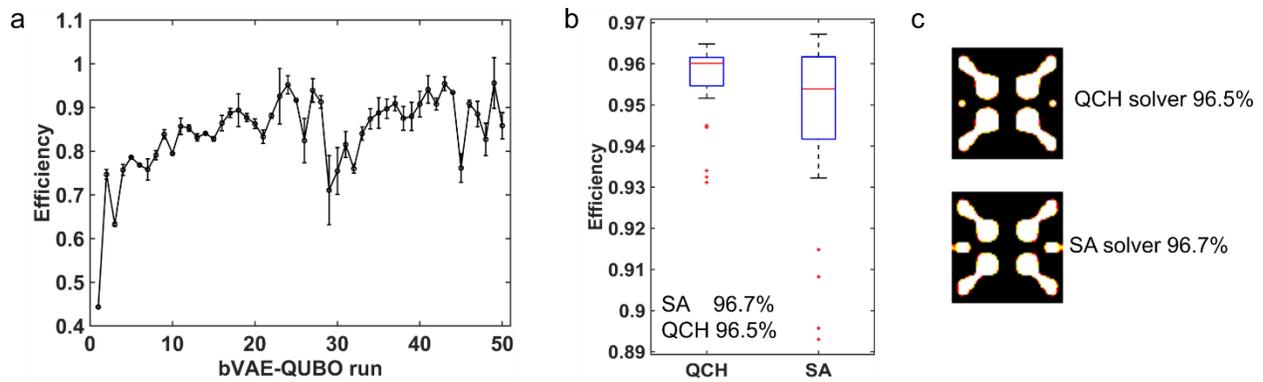

Figure 3. bVAE-QUBO assisted with a hybrid (quantum-classical) sampler for optimizing high-efficiency thermal emitter designs. (a) Convergence plot of 50 iterations of quantum-classical hybrid sampler assisted bVAE-QUBO. (b) Efficiency distribution comparison of the top 100 thermal emitter designs sampled via the hybrid (left) and simulated annealing (right) samplers. Box plot shows the median (red line), interquartile range (box) and outliers (red markers). Here, labeling indicates maximum efficiencies within each set. (c) Top view of the best thermal emitter designs sampled via the hybrid sampler (top) and simulated annealing (bottom) assisted bVAE-QUBO (white color corresponds to TiN, while black corresponds to air).

Quantum-Classical Hybrid Assisted bVAE-QUBO Framework

Along with the simulated annealer, we tested the bVAE-QUBO framework based on the quantum-classical hybrid sampler. The hybrid sampler is a high-quality server-side sampler hosted in the D-Wave Leap ecosystem that uses a mix of quantum annealing and classical sampling to sample from large QUBO's. Figure 3a shows the convergence plot of emitter efficiencies generated with hybrid sampling. As in the previous case, retraining the factorization machine with a refined dataset substantially increases the bVAE-QUBO framework's performance. We note that the main limitation of this approach is that the hybrid sampler returns one sample per bVAE-QUBO run. To augment the sampled dataset during the bVAE-QUBO run, we copied the sample 10 times and flipped a single bit for each copy. This allows us to expand the number of samples per epoch while sacrificing variance in the resulting designs. The comparison of the efficiency distributions obtained via the hybrid and simulated annealing-assisted bVAE-QUBO framework is shown in Fig. 3b. Here, we can see that using the hybrid sampler ensures narrower efficiency distribution with the median at 96% and interquartile range (25$^{th}$ to 75$^{th}$ percentile) between 95.5% and 96.2%. For the comparison, the efficiency distribution of the simulated annealing-based sampling has a 95.4% median and interquartile range between 94.2% and 96.2%. Both approaches provide almost identical maximum efficiencies, 96.5% (hybrid sampler) and 96.7% (simulated annealing). Corresponding thermal emitter designs are shown in Fig. 3c. Interestingly, both samplers lead to the designs with identical topologies, with slightly different lateral dimensions of the antenna components. Such narrow distribution of the sampled design efficiencies in the hybrid case might be a consequence of a better optimization search provided by the quantum annealing part of the sampler, which ensures a higher probability of locating an optimum in comparison with the classical simulated annealing algorithm.

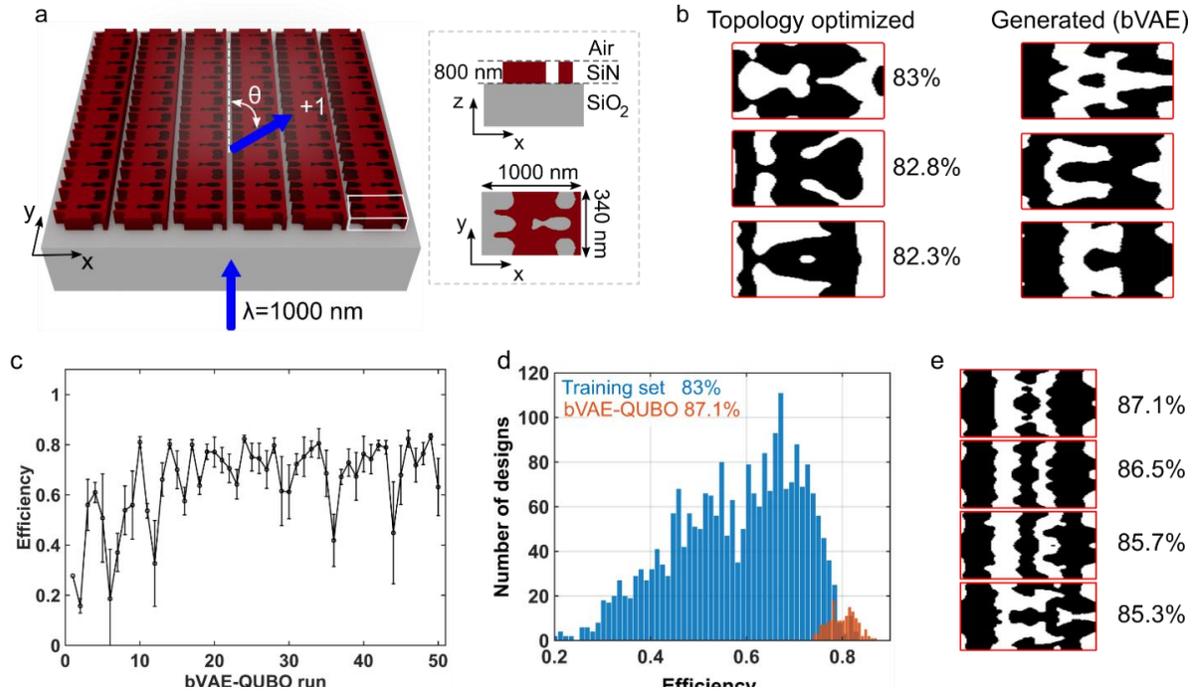

Figure 4. bVAE-QUBO based diffractive meta-grating optimization. (a) Schematics of the meta-grating optimization domain. The main goal of topology optimization is to determine material distribution inside the optimization region (highlighted by white box) placed on the silicon dioxide substrate that ensures highest possible deflection efficiency at $\theta$ deflection angle. The inset shows the configuration of the unit cell. (b) Best designs in the topology optimized training set and examples of bVAE sampled metagrating designs (white color - SiN, black - air). (c) Convergence plot for 50 runs of the simulated annealing-assisted bVAE-QUBO framework. (d) Efficiency distributions of the training set used for bVAE training (2000 designs, blue histogram) and 100 designs generated via the bVAE-QUBO framework (orange). (e) Designs of best designs sampled by the bVAE-QUBO.

### Inverse Design of Diffractive Meta-Gratings.

In the second case study, we optimize dielectric, free-form diffractive gratings for beam steering. Different types of dielectric meta-structures, metasurfaces, and meta-gratings have been used for various imaging applications[65], spectroscopy[66], as well as integrated optics applications[67–69]. The development of the dielectric antenna designs is one of the major steps in the meta-structure design pipeline. It has recently been demonstrated that advanced optimization frameworks, such as genetic algorithms[70] and adjoint topology optimization[71], can be used to develop various types of dielectric meta-devices. We apply the bVAE-QUBO framework to optimize silicon nitride (SiN) meta-gratings for deflecting normally incident light to a pre-defined angle $\theta$. The main goal of the optimization is to determine binary (SiN and air) material distribution within the optimization region, which maximizes the transmitted energy of the normally incident plane wave into +1 diffraction order at a 60-degree deflection angle (Fig. 4a).

Training the bVAE network On Topology Optimized Meta-grating Designs.

Adjoint topology optimization is used to generate 2000 SiN freeform meta-gratings (Supplementary Materials, Section S2)[71]. Figure 4b shows the best three designs in the training set. The bVAE network is trained on 100×100 pixelated patterns to compress the continuous-space representation of meta-grating into 500-dimensional binary compressed space. Some of the generated meta-grating designs by the trained bVAE network are shown in Fig. 4b. The figure of merit labels are assessed with S4, a rigorous coupled-wave analysis (RCWA) solver[72,73]. The evolution of the efficiencies of the meta-grating sampled by the bVAE-QUBO framework is shown in Fig. 4c. Similar to the previous example, gradual refinement of the QUBO parameters through retraining of the factorization machine significantly increases efficiencies of sampled designs. Within each iteration, the bVAE-QUBO samples ~10 meta-grating designs, generating 500 designs in total. Figure 4d shows a comparison of efficiencies of the bVAE training set (2000 topology optimized designs, blue bars) and the most efficient 100 designs sampled with the simulated annealing-assisted bVAE-QUBO framework (orange bars). The best designs in the topology optimized training set ensure 83% beam deflection efficiency, while the best design in the bVAE-QUBO set ensures 87.1 %. Figure 4e shows the meta-grating designs of the four highest efficiency designs sampled via the bVAE-QUBO framework. The figure indicates that similar to the previous case study, the bVAE-QUBO framework samples high-efficiency designs and produces significantly better designs than conventional topology optimization.

With regard to runtime performance, we highlight time requirements for each of the bVAE-QUBO's steps in our case studies. Training the bVAE network takes ~20 min using PyTorch, while training of the VGGnet regression model requires ~24 min on a standard desktop (Intel Core i7, 2.8 GHz CPU, 16 GB, Nvidia GeForce GTX 1050 GPU). Both the factorization machine training and execution of the bVAE-QUBO sampling are realized on the Google Colab platform. While the simulated annealing-assisted bVAE-QUBO requires only 3 s to sample at least one design, the hybrid sampler has a preset minimum annealing time of 10 s required to sample one design. The main bottlenecks come from the QUBO sampling time and determining the figure of merit of the device. More details are available in the Supplementary Materials, Section S3.

**Conclusion**

Within this work, we developed a unique framework that maps a broad range of continuous optimization problems into QUBO problems, which can be optimized by any available QUBO sampler. The developed binary variational autoencoder assisted QUBO framework reformulates a continuous-space problem into a QUBO problem and maps the constructed binary, compressed space into a QUBO sampler through a factorization machine model. The performance of the developed technique is demonstrated on two case studies of inverse design problems in nanophotonics, (i) thermal emitter topologies for TPV applications and (ii) diffractive meta-gratings for high-efficiency beam steering. This work is inspired by a recent, factorization machine-assisted QUBO framework applied for optimization of "checkerboard" type multi-layer metamaterial structures by setting "one to one" mapping of material pixels of the structure into the system of qubits of the D-Wave machine[21]. In contrast to [21], the bVAE-QUBO framework can (i) map continuous-space optimization problems without imposing those "checkerboard"-type solutions in the problem structure, and hence, (ii) reduce the dimension of the parametric space of the continuous domain by constructing a compressed binary space representation. Such generic formalism of the bVAE-QUBO framework opens up the possibility for mapping a broad range of highly constrained optimization problems of optics, chemistry, mechanics, finance, and computer science into any available QUBO sampler. While the current study showcased the performance of the bVAE-QUBO-based framework on classical and quantum-classical hybrid samplers, future work will focus on the realization of the bVAE-QUBO with a full quantum annealer or universal quantum computing. This work can be extended to similar frameworks with general, non-stoquastic Hamiltonians for adiabatic optimization.

**Acknowledgments**

This work is supported by the National Science Foundation award 2029553-ECCS, DARPA/DSO Extreme Optics and Imaging (EXTREME) Program (HR00111720032), and the U.S. Department of Energy (DOE), Office of Science through the Quantum Science Center (QSC), a National Quantum Information Science Research Center. We want to acknowledge and thank Oak-Ridge National Lab and the NASA Ames Center for allowing us to access their D-Wave 2000Q and D-Wave Advantage.

# References

1. Hoffman, K. L. Combinatorial optimization: Current successes and directions for the future. *J. Comput. Appl. Math.* **124**, 341–360 (2000).
2. Cook, S. A. The complexity of theorem-proving procedures. *Proceedings of the third annual ACM symposium on Theory of computing - STOC '71* 151–158 (1971). doi:10.1145/800157.805047
3. Kochenberger, G. *et al.* The unconstrained binary quadratic programming problem: a survey. *J. Comb. Optim.* **28**, 58–81 (2014).
4. Glover, F. *et al.* Quantum Bridge Analytics I: a tutorial on formulating and using QUBO models. *4OR* **17**, 335–371 (2019).
5. Jiang, S. *et al.* Quantum Annealing for Prime Factorization. *Sci. Rep.* **8**, 17667 (2018).
6. Xia, R. & Kais, S. Quantum machine learning for electronic structure calculations. *Nat. Commun.* **9**, 4195 (2018).
7. Laughhunn, D. J. Quadratic Binary Programming with Application to Capital-Budgeting Problems. *Oper. Res.* **18**, 454–461 (1970).
8. Boros, E. & Hammer, P. L. The max-cut problem and quadratic 0–1 optimization; polyhedral aspects, relaxations and bounds. *Ann. Oper. Res.* **33**, 151–180 (1991).
9. Kochenberger, G. A. *et al.* Solving large scale Max Cut problems via tabu search. *J. Heuristics* **19**, 565–571 (2013).
10. Wang, F. & Xu, Z. Metaheuristics for robust graph coloring. *J. Heuristics* **19**, 529–548 (2013).
11. Neukart, F. *et al.* Traffic Flow Optimization Using a Quantum Annealer. *Front. ICT* **4**, (2017).
12. Alidaee, B. *et al.* A new modeling and solution approach for the number partitioning problem. *J. Appl. Math. Decis. Sci.* **2005**, 113–121 (2005).
13. Lucas, A. Ising formulations of many NP problems. *Front. Phys.* **2**, (2013).
14. Finnila, A. B. *et al.* Quantum annealing: A new method for minimizing multidimensional functions. *Chem. Phys. Lett.* **219**, 343–348 (1994).
15. Das, A. & Chakrabarti, B. K. Colloquium: Quantum annealing and analog quantum computation. *Rev. Mod. Phys.* **80**, 1061–1081 (2008).
16. Johnson, M. W. *et al.* Quantum annealing with manufactured spins. *Nature* **473**, 194–198 (2011).
17. Apolloni, B. *et al.* Quantum stochastic optimization. *Stoch. Process. their Appl.* **33**, 233–244 (1989).
18. Kadowaki, T. & Nishimori, H. Quantum annealing in the transverse Ising model. *Phys. Rev. E* **58**, 5355–5363 (1998).
19. Aharonov, D. *et al.* Adiabatic Quantum Computation Is Equivalent to Standard Quantum

Computation. *SIAM Rev.* **50**, 755–787 (2008).

20. King, A. D. *et al.* Scaling advantage over path-integral Monte Carlo in quantum simulation of geometrically frustrated magnets. *Nat. Commun.* **12**, 1113 (2021).

21. Kitai, K. *et al.* Designing metamaterials with quantum annealing and factorization machines. *Phys. Rev. Res.* **2**, 013319 (2020).

22. Kairys, P. *et al.* Simulating the Shastry-Sutherland Ising Model Using Quantum Annealing. *PRX Quantum* **1**, 020320 (2020).

23. Perdomo-Ortiz, A. *et al.* Opportunities and challenges for quantum-assisted machine learning in near-term quantum computers. *Quantum Sci. Technol.* **3**, 030502 (2018).

24. Preskill, J. Quantum Computing in the NISQ era and beyond. *Quantum* **2**, 79 (2018).

25. Matsubara, S. *et al.* Ising-Model Optimizer with Parallel-Trial Bit-Sieve Engine. *Advances in Intelligent Systems and Computing* 432–438 (2018). doi:10.1007/978-3-319-61566-0_39

26. Tsukamoto, S. *et al.* An accelerator architecture for combinatorial optimization problems. *Fujitsu Sci. Tech. J.* **53**, 8–13 (2017).

27. Hastings, M. B. The Power of Adiabatic Quantum Computation with No Sign Problem. *arXiv* (2020).

28. Rendle, S. Factorization machines. *Proceedings of the2010 IEEE International Conference on Data Mining* 995–1000 (2010).

29. Pu, Y. *et al.* Variational Autoencoder for Deep Learning of Images, Labels and Captions. *Adv. Neural Inf. Process. Syst.* 2352–2360 (2016).

30. Kingma, D. P. & Welling, M. Auto-Encoding Variational Bayes. *2nd Int. Conf. Learn. Represent. ICLR 2014 - Conf. Track Proc.* (2013).

31. Jang, E. *et al.* Categorical Reparameterization with Gumbel-Softmax. *5th Int. Conf. Learn. Represent. ICLR 2017 - Conf. Track Proc.* (2016).

32. Maddison, C. J. *et al.* The Concrete Distribution: A Continuous Relaxation of Discrete Random Variables. *5th Int. Conf. Learn. Represent. ICLR 2017 - Conf. Track Proc.* (2016).

33. Sicks, R. *et al.* A lower bound for the ELBO of the Bernoulli Variational Autoencoder. *arXiv* (2020).

34. Dattani, N. *et al.* Pegasus: The second connectivity graph for large-scale quantum annealing hardware. *arXiv* (2019).

35. King, J. *et al.* Benchmarking a quantum annealing processor with the time-to-target metric. (2015).

36. Christiansen, R. E. & Sigmund, O. A tutorial for inverse design in photonics by topology optimization. *arXiv* (2020).

37. Jensen, J. S. & Sigmund, O. Systematic design of photonic crystal structures using topology optimization: Low-loss waveguide bends. *Appl. Phys. Lett.* **84**, 2022–2024 (2004).

38. Molesky, S. *et al.* Inverse design in nanophotonics. *Nat. Photonics* **12**, 659–670 (2018).
39. Lalau-Keraly, C. M. *et al.* Adjoint shape optimization applied to electromagnetic design. *Opt. Express* **21**, 21693 (2013).
40. Lin, Z. *et al.* Topology optimization of freeform large-area metasurfaces. *Opt. Express* **27**, 15765 (2019).
41. Jafar-Zanjani, S. *et al.* Adaptive Genetic Algorithm for Optical Metasurfaces Design. *Sci. Rep.* **8**, 11040 (2018).
42. Zhu, D. Z. *et al.* Optimal High Efficiency 3D Plasmonic Metasurface Elements Revealed by Lazy Ants. *ACS Photonics* **6**, 2741–2748 (2019).
43. Ma, W. *et al.* Deep-Learning-Enabled On-Demand Design of Chiral Metamaterials. *ACS Nano* **12**, 6326–6334 (2018).
44. Peurifoy, J. *et al.* Nanophotonic particle simulation and inverse design using artificial neural networks. *Sci. Adv.* **4**, eaar4206 (2018).
45. Jin, L. *et al.* Dielectric multi-momentum meta-transformer in the visible. *Nat. Commun.* **10**, 4789 (2019).
46. Nadell, C. C. *et al.* Deep learning for accelerated all-dielectric metasurface design. *Opt. Express* **27**, 27523 (2019).
47. Kiarashinejad, Y. *et al.* Knowledge Discovery in Nanophotonics Using Geometric Deep Learning. *Adv. Intell. Syst.* **2**, 1900132 (2020).
48. Kiarashinejad, Y. *et al.* Deep learning approach based on dimensionality reduction for designing electromagnetic nanostructures. *npj Comput. Mater.* **6**, 12 (2020).
49. Sajedian, I. *et al.* Finding the optical properties of plasmonic structures by image processing using a combination of convolutional neural networks and recurrent neural networks. *Microsystems Nanoeng.* **5**, 27 (2019).
50. Liu, Z. *et al.* Compounding Meta-Atoms into Metamolecules with Hybrid Artificial Intelligence Techniques. *Adv. Mater.* **32**, 1904790 (2020).
51. Ma, W. *et al.* Deep learning for the design of photonic structures. *Nat. Photonics* **15**, 77–90 (2021).
52. Jiang, J. *et al.* Deep neural networks for the evaluation and design of photonic devices. *Nat. Rev. Mater.* (2020). doi:10.1038/s41578-020-00260-1
53. Jiang, J. *et al.* Free-Form Diffractive Metagrating Design Based on Generative Adversarial Networks. *ACS Nano* **13**, 8872–8878 (2019).
54. Jiang, J. & Fan, J. A. Global Optimization of Dielectric Metasurfaces Using a Physics-Driven Neural Network. *Nano Lett.* **19**, 5366–5372 (2019).
55. Kudyshev, Z. A. *et al.* Machine-learning-assisted metasurface design for high-efficiency thermal emitter optimization. *Appl. Phys. Rev.* **7**, 021407 (2020).
56. Kudyshev, Z. A. *et al.* Machine learning assisted global optimization of photonic devices.

*Nanophotonics* **10**, 371–383 (2020).

57. Lenert, A. *et al.* A nanophotonic solar thermophotovoltaic device. *Nat. Nanotechnol.* **9**, 126–130 (2014).

58. Bierman, D. M. *et al.* Enhanced photovoltaic energy conversion using thermally based spectral shaping. *Nat. Energy* **1**, (2016).

59. Yeng, Y. X. *et al.* Enabling high-temperature nanophotonics for energy applications. *Proc. Natl. Acad. Sci.* **109**, 2280–2285 (2012).

60. Reddy, H. *et al.* Temperature-Dependent Optical Properties of Plasmonic Titanium Nitride Thin Films. *ACS Photonics* **4**, 1413–1420 (2017).

61. Chirumamilla, M. *et al.* Large-Area Ultrabroadband Absorber for Solar Thermophotovoltaics Based on 3D Titanium Nitride Nanopillars. *Adv. Opt. Mater.* **5**, 1700552 (2017).

62. Gui, L. *et al.* Nonlinear Refractory Plasmonics with Titanium Nitride Nanoantennas. *Nano Lett.* **16**, 5708–5713 (2016).

63. Ding, F. *et al.* A review of gap-surface plasmon metasurfaces: fundamentals and applications. *Nanophotonics* **7**, 1129–1156 (2018).

64. Hsiao, H.-H. *et al.* Fundamentals and Applications of Metasurfaces. *Small Methods* **1**, 1600064 (2017).

65. Genevet, P. & Capasso, F. Holographic optical metasurfaces: a review of current progress. *Reports Prog. Phys.* **78**, 024401 (2015).

66. Pors, A. *et al.* Plasmonic metagratings for simultaneous determination of Stokes parameters. *Optica* **2**, 716 (2015).

67. Chang-Hasnain, C. J. & Yang, W. High-contrast gratings for integrated optoelectronics. *Adv. Opt. Photonics* **4**, 379 (2012).

68. Quevedo-Teruel, O. *et al.* Roadmap on metasurfaces. *J. Opt.* **21**, 073002 (2019).

69. Yu, N. & Capasso, F. *Flat optics with designer metasurfaces. Nature materials* **13**, 139–50 (2014).

70. Campbell, S. D. *et al.* Review of numerical optimization techniques for meta-device design [Invited]. *Opt. Mater. Express* **9**, 1842 (2019).

71. Sell, D. *et al.* Large-Angle, Multifunctional Metagratings Based on Freeform Multimode Geometries. *Nano Lett.* **17**, 3752–3757 (2017).

72. Liu, V. & Fan, S. S4 : A free electromagnetic solver for layered periodic structures. *Comput. Phys. Commun.* **183**, 2233–2244 (2012).

73. Roberts, C. M. *et al.* Diffractive interface theory: nonlocal susceptibility approach to the optics of metasurfaces. *Opt. Express* **23**, 2764 (2015).

# Supplementary materials

# for

# A Machine Learning Framework for Quantum Sampling of Highly-Constrained, Continuous Optimization Problems

Blake A. Wilson[1,3,*], Zhaxylyk A. Kudyshev[1,3,*], Alexander V. Kildishev[1],
Sabre Kais[2,3], Vladimir M. Shalaev[1,3], and Alexandra Boltasseva[1,3]

[1]School of Electrical and Computer Engineering, Birck Nanotechnology Center and Purdue Quantum Science and Engineering Institute, Purdue University, West Lafayette, IN, USA
[2]School of Chemistry, Purdue University, West Lafayette, IN 47907, USA
[3]The Quantum Science Center (QSC), a National Quantum Information Science Research Center of the U.S. Department of Energy (DOE), Oak Ridge, TN 37931
[*]authors with equal contribution

## Section S.1. Training the Binary Variational Autoencoder and Factorization Machine

### S.1.1 Structure and Training of a Binary Variational Autoencoder

Within this work, we map the continuous design space found in many physically constrained problems into a binary design space. First, we construct an injective, invertible function, $g : \{0,1\}^n \to \mathbb{R}^{m \times m}$, that maps each binary vector, $x \in \{0,1\}^n$, in the domain to only one design for the problem of interest in a discretized space $\mathbb{R}^{m \times m}$. Constructing $g$ can be done via training of a binary variational autoencoder (bVAE). The bVAE is a deep neural network consisting of an encoder and a decoder. The encoder is a network with one input layer of $m \times m$ dimensions and two hidden dense layers with 512 and 256 neurons and ReLU activation functions. The decoder has an inverted structure to the encoder, two layers with 256 and 512 neurons, and one output layer. The structure of the network is shown in Fig. S1a. Specifically, the bVAE network learns how to compress continuous space designs into a binary, latent space and then reconstruct them. Naturally, the bVAE decoder acts as $g$, and the bVAE encoder acts as the inverse of $g$. The bVAE is trained by minimizing both the reconstruction loss for a design and the Kullback-Leibler divergence loss, $\mathcal{L}_{bVAE}$ [1]. The latter defines the deviation of the recognition distribution (obtained with the model data) from the pre-defined prior,

$$\mathcal{L}_{bVAE} = KL[q(z|y_m)|p(z)] - log[p(y_m|z)] \qquad (S1.1)$$

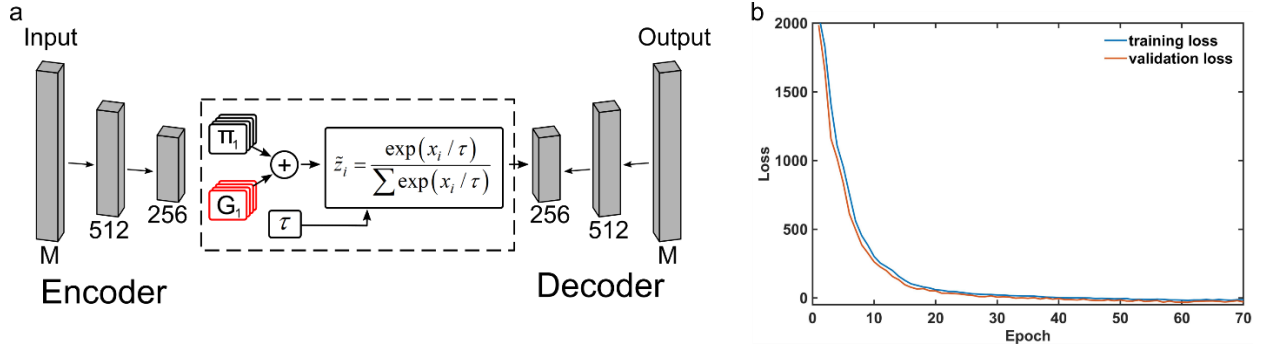

**Figure S1. Binary variational autoencoder training.** (a) Structure of the bVAE network and Gumbel-softmax re-parameterization of the binary latent variable. (b) Evolution of training and validation loss of bVAE network during training on thermal emitter design set.

The main difference between the bVAE and the typical VAE network is that the priors, p(z), and recognition model, q(z|y), are under different distributions. The bVAE is under a Bernoulli distribution, while the typical VAE's distribution is assumed to be Gaussian. The main problem with the bVAE, like with the VAE, is that a latent variable $x$ needs to be stochastically sampled with a pre-defined distribution to properly backpropagate the error for training the stochastic nodes in the network. This can be circumvented if we express the sample $x \sim p_\theta(z)$ such that the gradient can flow from the cost function to the set of the parameters $\theta$ (output of the encoder) without encountering stochastic nodes. For example, in a VAE network, the sampling of a latent variable with Gaussian distribution is realized by re-parameterization $x \sim N(\mu, \sigma)$ as $x = \mu + \sigma \cdot \varepsilon$, where $\varepsilon \sim N(0,1)$ and $(\mu, \sigma)$ are parameters of the encoder. This re-parameterization allows us to calculate their derivatives with respect to $\mu$ and $\sigma$ and use $\varepsilon$ as an additional input parameter sampled during each training epoch. We used the Gumbel-softmax re-parameterization trick to backpropagate the error in the bVAE, which is a similar re-parameterization to the standard VAE [1,2]. Specifically, Gumbel-softmax is a re-parameterization trick for a distribution that we can smoothly deform into the categorical distribution during the training process. Gumbel-softmax samples the latent space vectors $\widetilde{x_i}$ based on the class probabilities $\pi_1, \pi_2 \cdots \pi_n$ of the categorical representation as:

$$\widetilde{x_i} = \frac{\exp\left[\frac{\log(\pi_i) + G_i}{\tau}\right]}{\sum_{j=1}^{n} \exp\left[\frac{\log(\pi_j) + G_j}{\tau}\right]}, \quad i = 1, \ldots, n \qquad (S1.2)$$

here $G_i$ are independent and identically distributed variables sampled from Gumbel distribution **Gumbel(0,1)**. $\tau$ is a "temperature" parameter that controls how closely samples from Gumbel-softmax distribution approximates those from the true categorical distribution. During the training process, $\tau$ is gradually "annealed" from $\tau_{max}$ down to $\tau_{min}$, which is a good approximation to a categorical latent space distribution. We swept the $\tau$ parameter from

$\tau_{max} = 5$ to $\tau_{min} = 0.4$ with annealing rate $\gamma = 0.0003$. The evolution of the temperature follows an iterative form, $\tau_{\text{epoch}+1} = \tau_{\text{epoch}}\exp(-\gamma \cdot \text{epoch})$. We used a stochastic gradient descent optimization method, Adam (Adaptive Moment Estimation) [3], available through the Keras, and TensorFlow Python API during the training loop of the bVAE. The evolution of the training and validation losses of the bVAE network trained on 5000 thermal emitter designs are shown in Fig. S1b. 85% of the design set is used for training and 15% for validation.

### S.1.2 Pseudo-Boolean Structure of the Factorization Machine

Introduced by Rendle for learning sparse feature interactions, factorization machines are very useful, low-capacity models [4]. Consider a map, $h: \mathbb{R}^{m \times m} \to \mathbb{R}$, that calculates the figure of merit of a discretized, continuous space design. Then, $h(g): \{0,1\}^n \to \mathbb{R}$, where $g$ is the bVAE decoder, maps a binary vector in the compressed space of the bVAE to its figure of merit. Let $\tilde{y}(x) = h[g(x)]$, then $\tilde{y}(x)$ is a pseudo-boolean function. If we restrict the domain of the factorization machine to a binary space, then its model equation is isomorphic to an exhaustive pseudo-boolean function,

$$y(x) = \sum_{i=0}^{n} \left\langle v_i^{(1)} \right\rangle x_i + \sum_{i_1=0}^{n} \sum_{i_2 > i_1}^{n} \left\langle v_{i_1}^{(2)}, v_{i_2}^{(2)} \right\rangle x_{i_1} x_{i_2} + \sum_{i_1=0}^{n} \sum_{i_2 > i_1} \cdots \sum_{i_n > i_{n-1}} \left\langle v_{i_1}^{(2)}, \dots, v_{i_n}^{(n)} \right\rangle x_{i_1} \dots x_{i_n} \quad (S1.3)$$

For each polynomial degree, $d$, there exists a factorization matrix $v^{(d)}$. Given a factorization matrix $v^{(d)} \in \mathbb{R}^{n \times k}$, we can define a coefficient for the polynomial $x_{i_1} \dots x_{i_d}$ of degree $d$ as $\left\langle v_{i_1}^{(d)}, \dots, v_{i_d}^{(d)} \right\rangle$, which is a dot product between rows $i_1, i_2, \dots, i_d$ in $v^{(d)}$. The advantage of factorization machines is that their polynomial coefficients are determined by row interactions in their factorization matrix, which couples a change in one coefficient to a change in all the rows associated with that coefficient. This makes them a lower capacity model than a model where each polynomial coefficient is unique and decoupled from every other coefficient. However, this means that the model can infer coefficients under sparse training sets.

If a factorization machine is trained to approximate $\tilde{y}(x)$, then we can treat it as a surrogate model to $\tilde{y}(x)$ and sample its global optimum in place of $\tilde{y}(x)$, thereby sampling globally optimal designs within its highly compressed space. Naturally, this model's space complexity can be exponentially large with respect to the size of $x$ if $\tilde{y}(x)$ is a full pseudo-boolean function. We can make a calculated cut in the number of terms by noticing that given any polynomial of degree, $d$, the number of input strings where the coefficient of the polynomial contributes is $2^{n-d}$. So, we argue that the highest priority coefficients are the low order polynomials where the probabilities of any coefficient from first-order terms or second-order terms contributing to the output value are $\frac{2^{n-1}}{2^n} = \frac{1}{2}$ and $\frac{2^{n-2}}{2^n} = \frac{1}{4}$ respectively. Additionally, sampling a second-order factorization machine as a surrogate model is much more feasible because QUBO solvers can minimize second-order/quadratic pseudo-boolean functions. By restricting the factorization machine to first and second-order terms, its model equation becomes,

$$y(x) = \sum_{i=0}^{n} \langle v_i^{(1)} \rangle x_i + \sum_{i_1=0}^{n} \sum_{i_2 > i_1} \langle v_{i_1}^{(2)}, v_{i_2}^{(2)} \rangle x_{i_1} x_{i_2} \tag{S1.4}$$

A crucial benefit to restricting the factorization machine to a second-order model is that QUBO objective functions are only second-order polynomials. A QUBO sampler finds the minimum input string to a second-order pseudo-boolean function via classical or quantum sampling algorithms.

$$\underset{x \in \{0,1\}^n}{argmin} \sum_{i=0}^{n} Q_i x_i + \sum_{i_1=0}^{n} \sum_{i_2 > i_1} Q_{i_1,i_2} x_{i_1} x_{i_2} \tag{S1.5}$$

If we used a higher-order factorization machine, it would need to be quadratized to a second-order polynomial before being minimized by a QUBO sampler [5]. The scaling for this process can introduce an exponential number of variables or take an exponential amount of time with respect to the degree or input size of the QUBO. So, we restricted our framework to second-order factorization machines. Then we can directly map it to a QUBO without quadratization, where $Q_i = \langle v_i^{(1)} \rangle$ and $Q_{i_1,i_2} = \langle v_{i_1}^{(2)}, v_{i_2}^{(2)} \rangle$.

### S.1.3  Training the Factorization Machine

For both nanophotonic applications, we found that binary vectors of size 500, i.e., $x \in \{0,1\}^{500}$, and factorization matrices of size 500 by 40, i.e., $v^{(2)} \in \mathbb{R}^{500 \times 40}$, were sufficient to find good designs. For factorization machine training, we constructed a training set of 11250 unique vectors by randomly sampling from the binary, compressed space of the bVAE and assessing the performance of the design via a pre-trained convolutional neural network (thermal emitters) or a rigorous coupled-wave analysis (diffraction gratings). The evolution of the loss function of the factorization machine for the thermal emitter showcase example is shown in Fig. S2. It is also important to note for training that QUBO samplers minimize an objective function while we want to maximize the figure of merit for a problem. One can circumvent this issue by training the factorization machine on $c - \tilde{y}(x)$, where $\forall x \in \{0,1\}^n : c > \tilde{y}(x)$. Then, the minimization of the factorization machine corresponds to the maximization of $\tilde{y}(x)$.

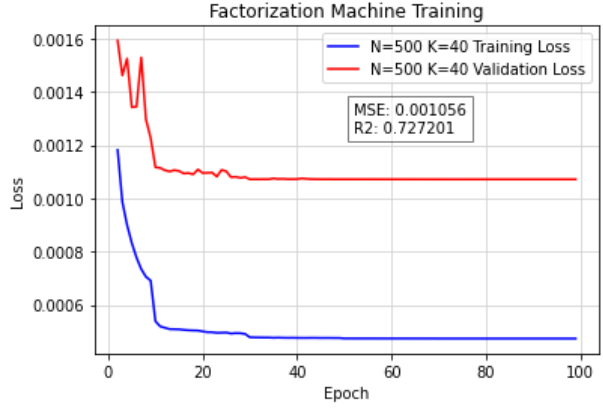

**Figure S2. Factorization machine training.** 100 epochs of training a second-order factorization machine-based regression model for the thermal emitter application. We found in practice that exceeding past 30 epochs did not improve the model's accuracy because it is a low-capacity model.

### S.1.4 Mapping a QUBO into an Ising Model

The restricted Ising Hamiltonian $H(\boldsymbol{\sigma})$ used by quantum annealers contains local biases $h$ and a quadratic term proportional to a qubit coupling matrix $J$ as:

$$H(\boldsymbol{\sigma}) = \sum_{i=1}^{N} h_i \sigma_i^{(z)} + \sum_{\langle i,j \rangle}^{N} J_{ij} \sigma_i^{(z)} \sigma_j^{(z)} \tag{S1.6}$$

$\sigma_i^{(z)}$ is a Pauli-Z matrix acting on the $i^{th}$ qubit's spin state. The spin state must collapse to an eigenvalue of -1 or +1 upon measurement. So, one can loosely consider this collapsed measurement value as a binary variable, $s_i \in \{-1, +1\}$.

$$H(\boldsymbol{s}) = \sum_{i=1}^{N} h_i s_i + \sum_{\langle i,j \rangle}^{N} J_{ij} s_i s_j \tag{S1.7}$$

Comparing (S1.5) and (S1.7), we can see the isomorphism between the parameters of the factorization machine and Ising model:

$$h_i \leftrightarrow w_i, \qquad J_{ij} \leftrightarrow \langle v_i, v_j \rangle \tag{S1.8}$$

However, the domain of the Ising Hamiltonian is the spin vectors, $\{-1, +1\}^n$, and the domain of the factorization machine is Boolean, $\{0,1\}^n$. There does exist a trivial transformation between the two domains, namely the invertible substitution $s_i = 2x_i - 1$. Hence, by training the factorization machine regression model and retrieving $\boldsymbol{w_0}, \boldsymbol{w}$, and coupling matrix $\langle \boldsymbol{v_i}, \boldsymbol{v_j} \rangle$, it is possible to construct an equivalent Ising or QUBO model.

S.1.5 Retrieving Optimal Vectors from a QUBO Sampler

Once the factorization machine equation is mapped into an equivalent QUBO or Ising form, we employ a QUBO sampler to find an input vector that minimizes the model's output. One thing to keep in mind is that the coupling matrix in a QUBO, $Q_{i,j}$, forms an undirected graph. Let $G = \{V, E\}$ be the connectivity graph for a QUBO coupling matrix, $Q$, where $V = \{1, 2, \ldots, n\}$ is the set of nodes in the graph, i.e. input bits, and $E = \{\{i,j\} \mid i, j \in V \text{ and } Q_{i,j} \neq 0 \text{ and } i \neq j\}$ is the set of edges in the graph. One would assume that the connectivity graph of a QUBO and its sampler must match to be compatible. However, this limitation only exists for QUBO samplers that use physical processes for sampling, such as quantum annealers. Additionally, minor-embedding techniques can convert $Q$ to another graph, $Q'$, that is compatible with the QUBO sampler and has the same minimum input vectors as $Q$. Unfortunately, this process introduces many more nodes and edges, and it is possible that the number of nodes and edges in $Q'$ exceeds that of the physical QUBO sampler.

A special case for this embedding is when considering fully connected graphs. If $Q$ is fully connected, such as with our factorization machine's QUBO, then the maximum number of nodes for which a $Q'$ exists for a given physical sampler is known as the "clique" size. The clique size for a QUBO sampler depends on the topology of the QUBO sampler and the number of available nodes. There are two available topologies for the D-Wave quantum annealer connectivity: Pegasus (D-Wave Advantage) [6] and Chimera (D-Wave 2000Q) [7]. The D-Wave Advantage quantum annealer has a Pegasus connectivity graph with a maximum clique size of 180 qubits, while the D-Wave 2000Q's Chimera graph clique size is 64 qubits. Along with quantum annealing, the D-Wave Leap ecosystem supports simulated annealing and quantum-classical hybrid samplers, which can handle any clique size. Naturally, the quantum annealer may ensure asymptotic speed-up over classical computers with respect to the QUBO size.

Unfortunately, we found in practice that our factorization machines required 500-dimensional input vectors to be good models, which exceeds the maximum clique size of the D-Wave Advantage. It is also worth noting that this obstacle can be overcome by using the decomposition QUBO solver that divides large QUBO problems into sub-problems [8] or by additional regularization of the bVAE training process, which can adapt the distribution of the binary, compressed space to better match the factorization machine model. Due to the aforementioned restriction of maximum clique size, we resorted to using D-Wave's simulated annealer and quantum-classical hybrid sampler.

## Section S.2.  Section S2. Generating a Topology Optimized Training Set

### S.2.1  Thermal emitter.

Within this work, we have used a previously developed adversarial autoencoder-based optimization framework to generate the training set for the bVAE-QUBO algorithm. The adversarial autoencoder network is initially trained on 200 topology optimized designs of a three-layered gap-plasmon structure. 200 topology optimized designs have been enlarged via a data argumentation technique developed in [9].

The adversarial autoencoder consists of three coupled neural networks: the encoder, the decoder/generator, and the discriminator [10]. **The encoder** takes a 4096-dimensional input vector (that corresponds to a 64 × 64 binary design pattern). The input is fed into the first of two fully connected, hidden layers with 512 neurons each and a ReLU activation function on the output of both layers. One batch normalization layer is coupled to the second hidden layer. The output is a 15-neuron layer. **The decoder** has the same architecture as the encoder but with the reversed sequence. The decoder generates a 4096-element output vector based on 15-dimensional, binary input. For the output layer, we use the hyperbolic tangent activation function. **The discriminator** takes a 15-dimensional latent vector as an input and has one output neuron for binary classification (fake/real). Here we use two hidden linear layers with 512 and 256 neurons. The two hidden layers use a ReLU activation function, and the output layer uses a sigmoid function.

Once the adversarial autoencoder network is trained, the decoder generates an additional 5000 designs. Moreover, to avoid time-consuming full-wave analysis during the execution of the bVAE-QUBO, a VGGnet convolutional neural network is trained for rapid estimation of a thermal emitter's efficiency based on its design. The VGGnet takes 64 by 64 images of the design as an input and passes it through three hidden layers, consisting of convolutional layers with ReLU activation functions. Each hidden layer is followed by a max-pooling layer, which ensures the down-sampling of the feature maps. The stack of convolutional layers is followed by one fully connected layer. The final layer has a linear activation function with a mean squared error loss function for efficiency prediction. The supervised training of the VGGnet is realized on the same 5000 designs. The VGGnet regression model ensures ~93% accuracy ($r^2$ coefficient) of predicting the design efficiency. Training the adversarial autoencoder and VGG networks is done similarly to [9].

### S.2.2 Diffraction Metagrating

For the diffraction metagrating example, we developed a topology optimized dataset by the adjoint topology optimization method. Here we follow the optimization framework developed in [11]. The main goal of the optimization is to determine binary material distribution (SiN in air) within the optimization region, which maximizes transmission into a +1-diffraction order at a 60-degree angle of a normally incident plane wave. The optimization region is set to be a **1 µm × 0.34 µm** region with a thickness of **0.8 µm** placed on top of the SiO$_2$ substrate. A **1 µm** wavelength plane wave excitation occurs from the substrate side. Topology optimization attempts to maximize the transmission efficiency of the incident light into a pre-defined diffraction order via maximization of the overlap integral between total field induced by the incident $[\mathbf{E^{fwd}}, \mathbf{H^{fwd}}]$ and the field induced by backward propagating adjoint field $[\mathbf{E^{bwd}}, \mathbf{H^{bwd}}]$. The overlap integral is calculated above the optimization region ($z = z_1$).

$$F = \left| \int_{y_{min}}^{y_{max}} \int_{x_{min}}^{x_{max}} \begin{pmatrix} \mathbf{E}^{fwd}(x,y,z_1) \times \mathbf{H}^{bwd}(x,y,z_1) - \\ \mathbf{E}^{bwd}(x,y,z_1) \times \mathbf{H}^{fwd}(x,y,z_1) \end{pmatrix} \cdot \mathbf{n}_z dx dy \right|^2 \qquad (S2.1)$$

The main goal of the adjoint formalism is to express the gradient $\partial F(x,y)/\partial \varepsilon$ as a function of field distributions inside the optimization region induced by forward and backward (adjoint) excitation. Such formalism obtains gradients at each location of the optimization region via only two full-wave analyses. More details on the adjoint topology optimization formalism for dielectric metagrating optimization can be found in [11].

### References
[1] E. Jang, S. Gu, and B. Poole, 5th Int. Conf. Learn. Represent. ICLR 2017 - Conf. Track Proc. (2016).
[2] C. J. Maddison, A. Mnih, and Y. W. Teh, 5th Int. Conf. Learn. Represent. ICLR 2017 - Conf. Track Proc. (2016).
[3] F. Chollet, Keras.Io (2015).
[4] S. Rendle, in *Proc. The2010 IEEE Int. Conf. Data Min.* (2010), pp. 995–1000.
[5] N. Dattani, ArXiv (2019).
[6] N. Dattani, S. Szalay, and N. Chancellor, ArXiv (2019).
[7] J. King, S. Yarkoni, M. M. Nevisi, J. P. Hilton, and C. C. McGeoch, (2015).
[8] M. Booth, S. P. Reinhardt, and A. Roy, Quantum Exec. Tech. Rep. (2017).
[9] Z. A. Kudyshev, A. V. Kildishev, V. M. . Shalaev, and A. Boltasseva, Appl. Phys. Rev. **7**, 021407 (2020).
[10] A. Makhzani, J. Shlens, N. Jaitly, I. Goodfellow, and B. Frey, arXiv:1511.05644 (2015).
[11] D. Sell, J. Yang, S. Doshay, R. Yang, and J. A. Fan, Nano Lett. **17**, 3752 (2017).